\documentclass[a4paper,fleqn,usenatbib,letters]{mnras}

\usepackage{newtxtext,newtxmath}
\usepackage[T1]{fontenc}
\usepackage{ae,aecompl}

\usepackage{graphicx}	
\usepackage{amsmath}	
\usepackage{amssymb}	
\usepackage{color}

\title[Novel modelling of UCXBs]{Novel modelling of ultra-compact X-ray binary evolution\\-- stable mass transfer from white dwarfs to neutron stars}

\author[Sengar et~al.]
{Rahul Sengar$^{1}$\thanks{E-mail: rsengar@astro.uni-bonn.de},
Thomas M. Tauris$^{2,1}$,
Norbert Langer$^{1}$,
Alina G. Istrate$^{3}$
\\
$^{1}$ Argelander-Institut f\"ur Astronomie, Universit\"at Bonn, Auf dem H\"ugel 71, D-53121 Bonn, Germany\\
$^{2}$ Max-Planck-Institut f\"ur Radioastronomie, Auf dem H\"ugel 69, D-53121 Bonn, Germany\\
$^{3}$ Center for Gravitation, Cosmology, and Astrophysics, Dept. of Physics, Uni. of Wisconsin-Milwaukee, Milwaukee, WI, 53201, USA
}

\date{Accepted 2017 April 26. Received 2017 April 26; in original form 2017 March 7}

\pubyear{2017}

\begin{document}
\label{firstpage}
\pagerange{\pageref{firstpage}--\pageref{lastpage}}
\maketitle

\begin{abstract}
Tight binaries of helium white dwarfs (He~WDs) orbiting millisecond pulsars (MSPs) will eventually ``merge'' due to gravitational damping of the orbit. The outcome has been predicted to be the production of long-lived ultra-compact X-ray binaries (UCXBs), in which the WD transfers material to the accreting neutron star (NS). 
Here we present complete numerical computations, for the first time, of such stable mass transfer from a He~WD to a NS.
We have calculated a number of complete binary stellar evolution tracks, starting from pre-LMXB systems, and evolved these to detached MSP+WD systems and further on to UCXBs. The minimum orbital period is found to be as short as 5.6~minutes. We followed the subsequent widening of the systems until the donor stars become  planets with a mass of $\sim 0.005\;M_{\odot}$ after roughly a Hubble time.
Our models are able to explain the properties of observed UCXBs with high helium abundances and we can identify these sources on the ascending or descending branch in a diagram displaying mass-transfer rate vs. orbital period. 
\end{abstract}

\begin{keywords}
binaries: close --- X-rays: binaries --- stars: mass-loss --- stars: neutron --- white dwarfs --- pulsars: general
\end{keywords}


\section{Introduction}

The detection of radio millisecond pulsars (MSPs) in close orbits with helium white dwarfs (He~WDs) raises interesting questions about their future destiny. 
One example is PSR~J0348+0432 \citep{afw+13} which has an orbital period of 2.46~hr. Due to continuous emission of gravitational waves, this system will ``merge'' in about 400~Myr. 
However, rather than resulting in a catastrophic event, once the WD fills its Roche lobe, the outcome is expected to be a long-lived ultra-compact X-ray binary \citep[UCXB,][]{webbink1979,nrj86,prp02,nyvt10,vnv+12,hie+13}. 
These sources are tight low-mass X-ray binaries (LMXBs) observed with an accreting neutron star (NS) and a typical orbital period of less than 60~min. 
Because of the compactness of UCXBs, the donor stars are constrained to be either a WD, a semi-degenerate dwarf or a helium star \citep{rjw82}.

Depending on the mass-transfer rate, the UCXBs are classified in two categories: persistent and transient sources. Until now, only 14 UCXBs have been confirmed (9 persistent, 5 transient), and an additional 14 candidates are known. 
Therefore, we can infer that UCXBs are difficult to detect or represent a rare population. Earlier studies \citep[e.g.][]{itl14} have suggested the need for extreme fine tuning of initial parameters (stellar mass and orbital period of the LMXB progenitor systems) in order to produce an UCXB from an LMXB system.
Analytical investigations by \citet{vnv+12} and \citet{vnvj12b} on the evolution of UCXBs with NS or black hole accretors reveal that these systems can evolve to orbital period of $100-110\:{\rm min}$, thereby explaining the existence of the so-called diamond planet pulsar \citep{bbb+11}.

UCXBs are detected with different chemical compositions in the spectra of their accretion discs \citep[e.g. H, He, C, O and Ne,][]{nyvt10}. To explain this diversity requires donor stars which have evolved to different levels of nuclear 
burning and interior degeneracy, and therefore to different scenarios for the formation of UCXBs.
Since a large fraction of the UCXBs are found in globular clusters, some of these UCXB systems could also have formed via stellar exchange interactions \citep{fpr75}. 
For a $1.4\;M_{\odot}$ NS accretor, only CO~WDs with a mass of $\la 0.4\;M_{\odot}$ lead to stable UCXB configurations \citep{vnv+12}, although recent hydrodynamical simulations suggest that this critical WD mass limit could be lower \citep{bdc17}.

Here we focus on numerical computations covering, for the first time, complete evolution of NS-main sequence star binaries which evolve into LMXBs and later produce UCXBs with a He~WD donor star. The evolution is terminated when the donor reaches a mass of $\sim\!0.005\;M_{\odot}$ (about 5 Jupiter masses) after several Gyr, with a radius close to the maximum radius of a cold planet.

In Section~\ref{sec:setup}, we present the applied stellar evolution code, as well as key assumptions on binary interactions and a summary of our applied model. The results of our calculations are presented in Section~\ref{sec:results}. A comparison with the observed UCXBs systems is give in Section~\ref{sec:obs}, and finally we further discuss and summarise our results in Section~\ref{sec:summary}.
\section{Numerical code and initial setup}\label{sec:setup}
We applied the MESA code \citep[Modules for Experiments in Stellar Astrophysics,][and references therein]{pms+15} for calculating the  evolution of NS--main sequence star binaries. 
Our initial binary system consists of a $1.3\;M_{\odot}$ NS (treated as a point mass) and a zero-age main-sequence (ZAMS) donor star of mass $M_2=1.4\;M_{\odot}$ with solar chemical composition ($X=0.70$, $Y=0.28$ and $Z=0.02$). We investigated a range of initial orbital periods of $P_{\rm orb,i}\simeq 2-5\;{\rm days}$, with a total of 40 models. 
These orbits were assumed to be circular and synchronized.
We assumed standard loss of orbital angular momentum due to magnetic braking (only significant during the LMXB phase), gravitational wave (GW) radiation and mass loss \citep[e.g.][]{tv06}. We modelled the latter via Roche-lobe overflow (RLO) according to the isotropic re-emission model \citep{vdh94,ts99}, in which matter flows from the donor star to the accreting NS in a conservative manner and a fraction \citep[here $\beta =0.70$, see discussions in][]{itl14} is ejected from the vicinity of the NS, and at all times ensuring sub-Eddington mass-accretion rates ($|\dot{M}_2|<\dot{M}_{\rm Edd}\simeq 3.0\times 10^{-8}\;M_{\odot}\;{\rm yr}^{-1}$).
For the initial phase of the mass transfer in the UCXB stage (once the He~WD remnant initiates RLO to the NS), the mass-transfer rate is super-Eddington, and thus the NS accretion rate is limited to $\dot{M}_{\rm Edd}$. For simplicity, and to avoid too many free parameters, we do not include the possibility of a circumbinary (CB) disc \citep{sl12}, nor do we consider irradiation of the donor
star via pulsar winds or photons.

\section{Results}\label{sec:results}
\begin{figure}
\vspace{-0.7cm}
\begin{center}
\includegraphics[width=1.05\columnwidth,angle=0]{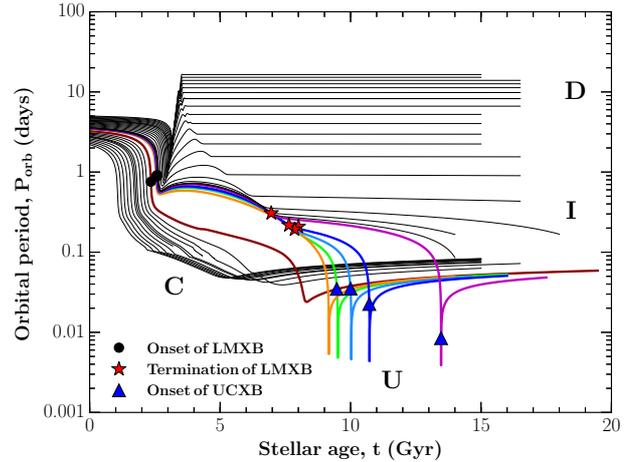}\\
\caption{Orbital period evolution of LMXBs. The outcome is either diverging (D) or intermediate (I) systems, converging (C) systems, or binaries (U) which detach, produce He~WDs and evolve into UCXBs. See also fig.~2 in \citet{itl14}. The colour codings of the tracks are identical to those in Figs.~\ref{fig:tracks}--\ref{fig:donor}.}
\label{fig:LMXB}
\end{center}
\end{figure}

\subsection{LMXB/pre-UCXB evolution}\label{subsec:preUCXB}
Fig.~\ref{fig:LMXB} shows the orbital period evolution as a function of age for several LMXBs with a range of $P_{\rm orb,i}=2.2-5.0\;{\rm days}$ and magnetic braking index $\gamma=5$ \citep{itl14}. 
For these close-orbit binaries, magnetic braking efficiently shrinks the orbits such that the companion star is forced to initiate RLO within $1-3\;{\rm Gyr}$. 
Based on the classification of \citet{ps88} and \citet{itl14}, we divide the orbital evolution of the resulting LMXBs into diverging, intermediate and converging systems. Notably, we find a narrow range of $P_{\rm orb,i}$ for which LMXB systems detach and produce a He~WD in a tight orbit \citep[marked by U, see also][for discussions]{itl14}. Such systems are observable as radio MSP binaries with typical $P_{\rm orb}\simeq 2-9\;{\rm hr}$ \citep[e.g. PSRs~J0751+1807 and J0348+0432,][]{elc97,afw+13}.
These MSP binaries will shrink their orbits further by GW radiation; most of them to the extent that their He~WDs are forced to fill their Roche~lobe within a total age of less than a Hubble time. At this stage the systems become UCXBs, typically when $P_{\rm orb}\simeq 10-50\;{\rm min}$, cf. Fig.~\ref{fig:tracks}. 
As a result of the ($M_{\rm WD},P_{\rm orb}$)--relation for He~WDs \citep{ts99,itl14} all our UCXBs initially have $M_{\rm WD}\simeq 0.15-0.17\;M_{\odot}$.

\subsection{UCXB evolutionary tracks}\label{subsec:tracks}
\begin{figure*}
\vspace{-0.5cm}
\begin{center}
\includegraphics[width=1.5\columnwidth,angle=0]{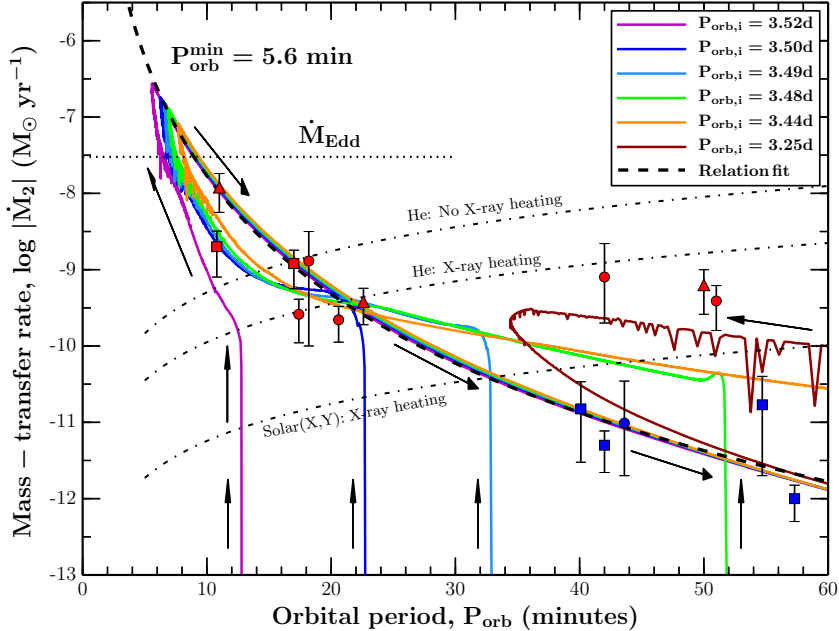}
\caption{Evolutionary tracks of our modelling of the UCXB phase in the ($P_{\rm orb},\,\dot{M}_2$)--plane, calculated for six different values of $P_{\rm orb,i}$, with RLO from the He~WD, following the detached epoch of an MSP--WD system -- except for the systems with $P_{\rm orb,i}=3.25$ and $3.44\;{\rm days}$ which never detach completely in their transition from an LMXB to an UCXB. 
The UCXB evolution (see black arrows) begins on the vertical tracks, when the He~WD initiates RLO, and decreases $P_{\rm orb}$ due to GWs while climbing up the {\it ascending} branch until the tip of the track at $P_{\rm orb}^{\rm min}$, before the system settles on the common {\it declining} branch (dashed line) while $P_{\rm orb}$ steadily increases on a Gyr timescale. 
More details of $\dot{M}_2$ and the donor stars are found in Figs.~\ref{fig:mass-transfer-rate} and \ref{fig:donor}.
The observed UCXBs \citep{hie+13} are shown with red (persistent source) and blue (transient source) symbols. UCXBs marked with squares, triangles and circles correspond to systems with no information of spectra, with helium and without helium lines, respectively. The dot-dashed lines indicate accretion disk instability thresholds, depending on X-ray heating and chemical composition. The dotted line is $\dot{M}_{\rm Edd}$ for a NS accreting helium.}
\label{fig:tracks}
\end{center}
\end{figure*}

In Fig.~\ref{fig:tracks}, we plot evolutionary tracks for the UCXB phase (i.e. post-LMXBs) of six systems with values of $P_{\rm orb,i}$ between 3.25 and 3.52~days, cf. coloured tracks in Fig.~\ref{fig:LMXB}. 
Several features of UCXB formation can be seen from  Fig.~\ref{fig:tracks} which we now discuss in more detail.

Firstly, it is evident that LMXBs with $P_{\rm orb,i}$ below a certain threshold value (depending on initial values of $M_2$, $M_{\rm NS}$, chemical composition and treatment of magnetic braking; here $\sim\! 3.45\;{\rm days}$) will never detach from RLO to produce a He~WD. Their donor stars still possess a significant hydrogen content -- even in their cores -- and due to their very small nuclear burning rates they still have a mixture of hydrogen and helium when they finally become degenerate near the orbital period minimum, $P_{\rm orb}^{\rm min}\simeq 10-85\;{\rm min}$ \citep{ps81,rjw82,nrj86,prp02,vvp05}. 
These converging systems become hydrogen-rich UCXBs and are most likely the progenitor systems of the so-called black widow MSPs \citep[e.g.][]{esa98,bdh12,ccth13}.
For our two hydrogen-rich systems with $P_{\rm orb,i}=3.25\;{\rm days}$ (red track) and $P_{\rm orb,i}=3.44\;{\rm days}$ (orange track), we find $P_{\rm orb}^{\rm min}\simeq 35\;{\rm min}$ and 8~min, respectively.

Secondly, for UCXBs with He~WD donors, the larger the value of $P_{\rm orb,i}$, the smaller is $P_{\rm orb}$ at the onset of the UCXB phase. The reason is that in wider binaries He~WDs have larger masses \citep{ts99}; and, more important, since in wider systems it takes longer time for GWs to cause the He~WDs to fill their Roche lobe, they will be less bloated \citep{itla14}, i.e. more compact (and colder) by the time they reach the onset of the UCXB phase. 

\begin{figure}
\vspace{-0.4cm}
\begin{center}
\includegraphics[width=1\columnwidth,angle=0]{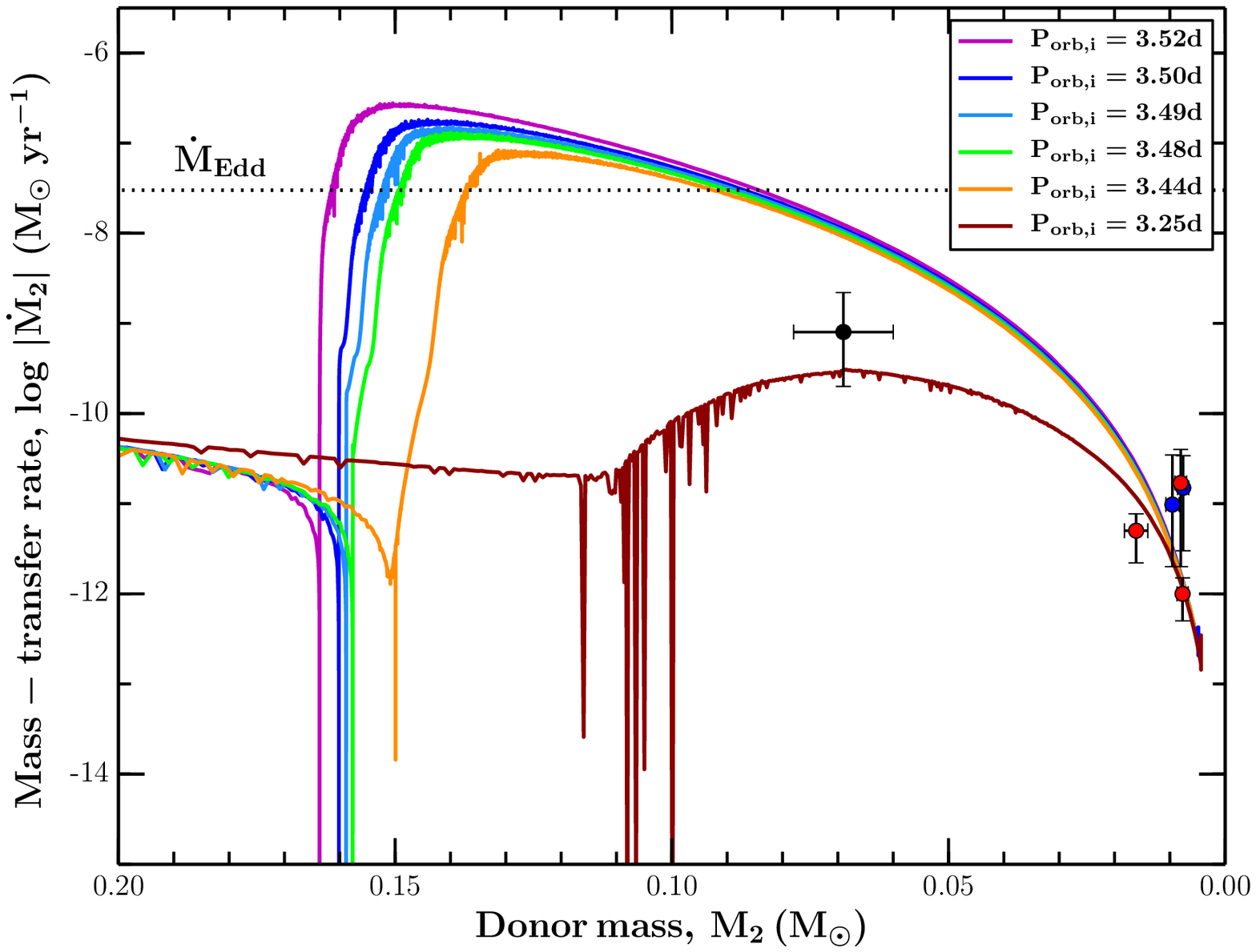}
\includegraphics[width=1.0\columnwidth,angle=0]{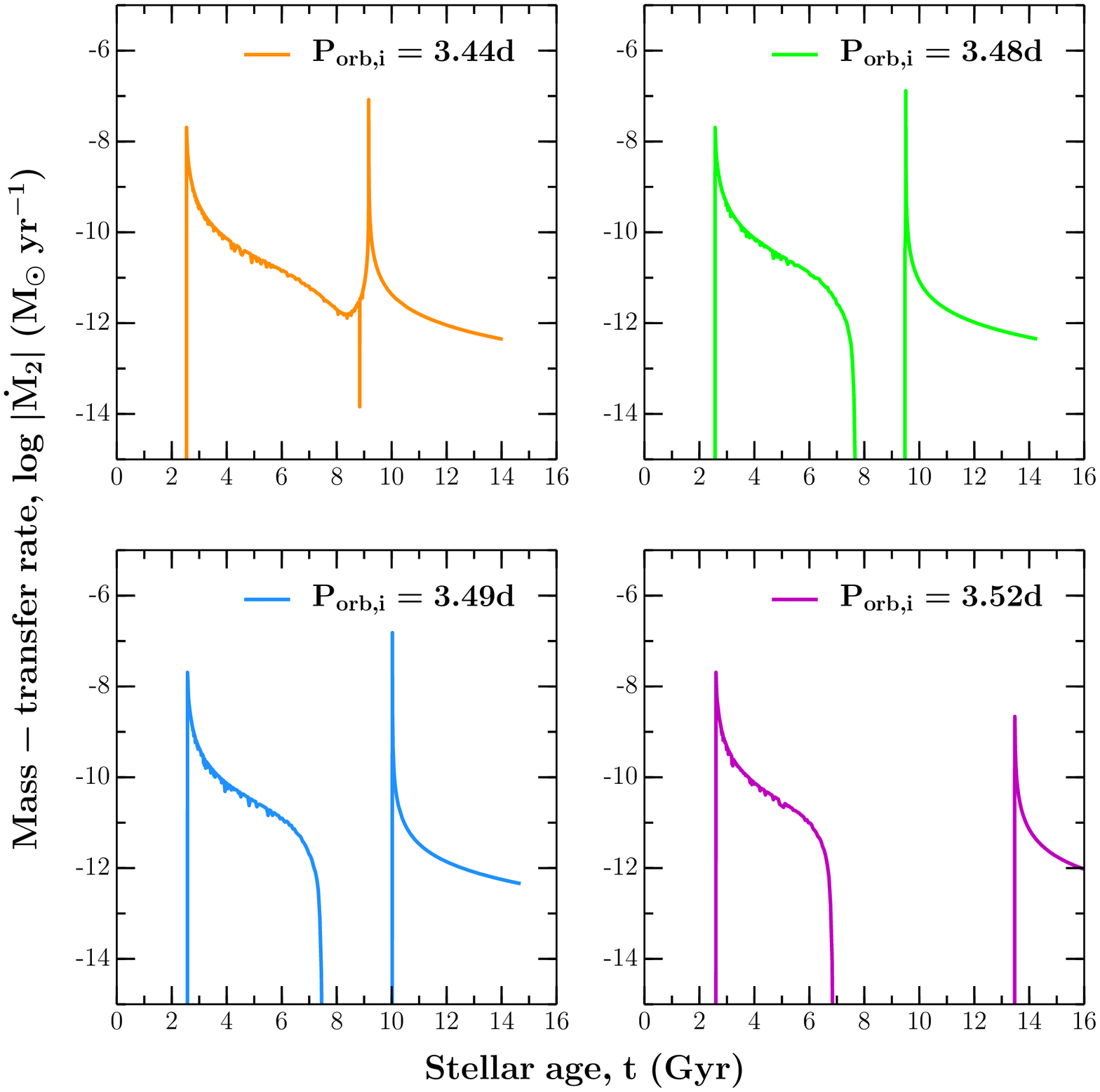}
\caption{Upper panel: mass-transfer rate, $|\dot{M}_2|$ as a function of donor star mass ($M_2$) for the UCXB tracks shown in Fig.~\ref{fig:tracks}. The observed WD donor masses of six known accreting X-ray MSP UCXBs \citep[five systems are transient,][]{pw12} are marked in red and blue colours which correspond to He and CO~WDs. The masses of three WD companions are very similar to one another and cluster at the same point. The WD marked in black colour could either be a WD or a helium star. Lower panels: four of the above tracks plotted in ($t,\,|\dot{M}_2|$)--diagrams.} 
\label{fig:mass-transfer-rate}
\end{center}
\end{figure}

Thirdly, we identify a unique pattern in the tracks of these UCXBs (see black arrows). They begin on the vertical tracks, when the He~WD initiates RLO, and continue decreasing $P_{\rm orb}$ due to GWs while climbing up the {\it ascending} branch until the tip of the track at $P_{\rm orb}^{\rm min}$. For our He~WD UCXBs we find typically 
$P_{\rm orb}^{\rm min}\simeq 5-7\;{\rm min}$. Following $P_{\rm orb}^{\rm min}$, which coincides with a maximum value of $|\dot{M}_2|\simeq 10\;\dot{M}_{\rm Edd}$ (see also Fig.~\ref{fig:mass-transfer-rate}), all systems settle on the common {\it declining} branch while $P_{\rm orb}$ steadily increases on a Gyr timescale, with the relation: 
$\log |\dot{M}_2/M_{\odot}\,yr^{-1}|=-5.15\cdot\log (P_{\rm orb}/{\rm min})-2.62$.

The shape of the UCXBs tracks can be understood from the ongoing competition between GW radiation and orbital expansion caused by mass transfer/loss.
The reason that the maximum value of $|\dot{M}_2|$ coincides with $P_{\rm orb}^{\rm min}$ is partly that the He~WD donor stars are fully degenerate, which means that their mass--radius exponent is negative (Fig.~4c), whereby they expand in response to mass loss \citep[at least after their residual hydrogen-rich envelope of a few $10^{-3}\;M_{\odot}$ has been removed,][]{kbs12}. 
The onset of RLO leads not only to very high mass-transfer rates (Fig.\ref{fig:mass-transfer-rate}) but also to an outward acceleration of the orbital size, as a result of the small mass ratio ($q\simeq 0.1$) between the two stars, such that at some point the rate of orbital expansion dominates over that of shrinking due to GW radiation. As the orbits widen further, the value of $|\dot{M}_2|$ decreases and the strength of GW radiation levels off due to its steep dependence on orbital separation and the systems settle on the common declining branch while the orbit expands at a continuously slower pace. 
An analogy of our described UCXB models can be made to RLO in double WD systems \citep{kbs12}. 

Our final $M-R$ tracks (Fig.~4c) terminating at 5 Jupiter masses ($\sim\!0.005\;M_{\odot}$) are in good agreement with (within 5\% of) the adiabatic helium models of \citet{db03} and the cold helium models of \citet{zs69}. For a comparison to $M-R$ tracks of cold planets, see Fig.~4d.
The He~WD donors never crystallise but remain Coulomb liquids with $\Gamma\le 35$. 
In all our models, the final mass of the (post) UCXB NS is $1.70\pm0.01\;M_{\odot}$, reflecting the assumed NS birth mass (here $1.3\;M_{\odot}$) and the accretion efficiency.

\begin{figure*}
\vspace{-0.5cm}
\begin{center}
\includegraphics[width=0.92\columnwidth,angle=0]{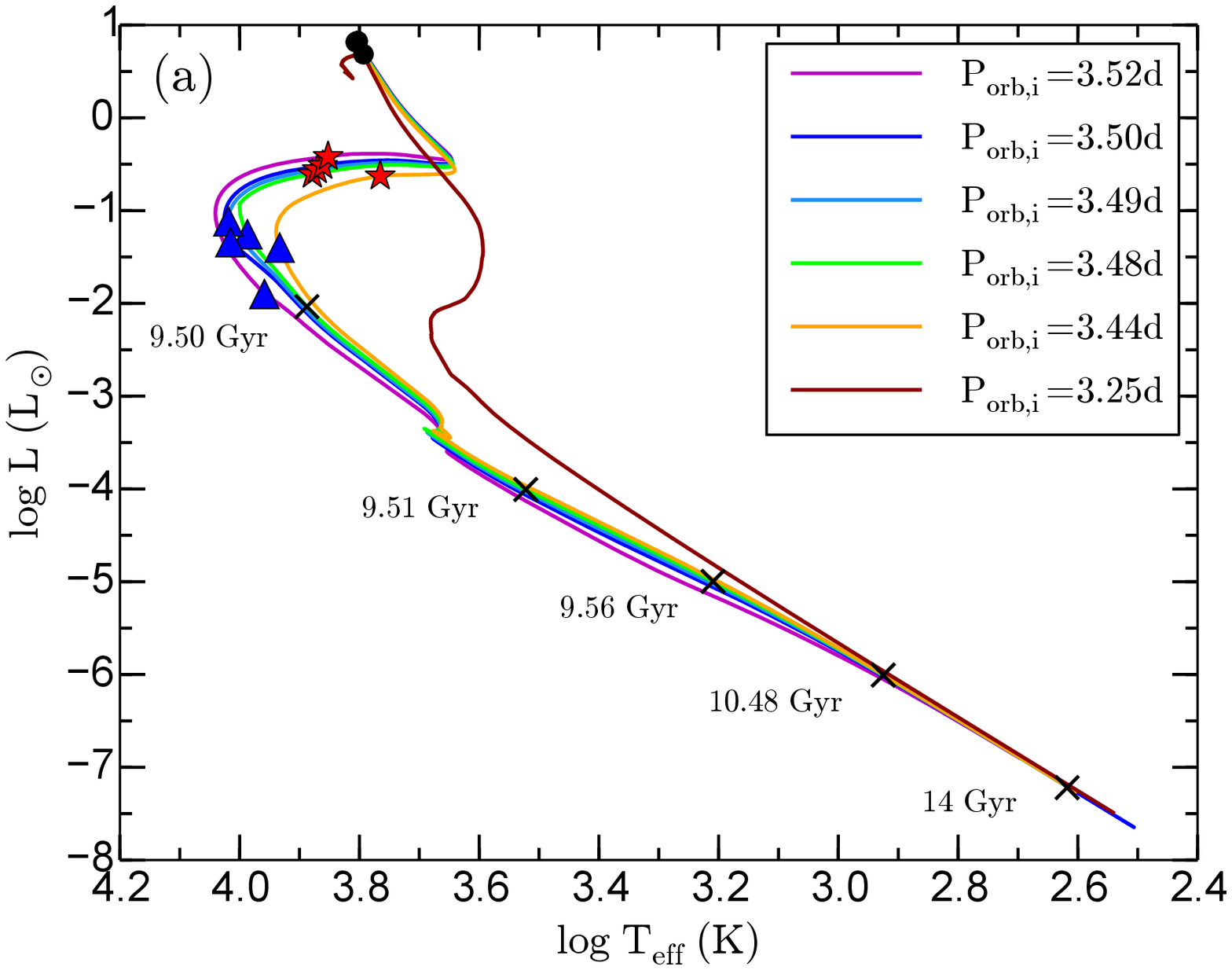}
\includegraphics[width=0.92\columnwidth,angle=0]{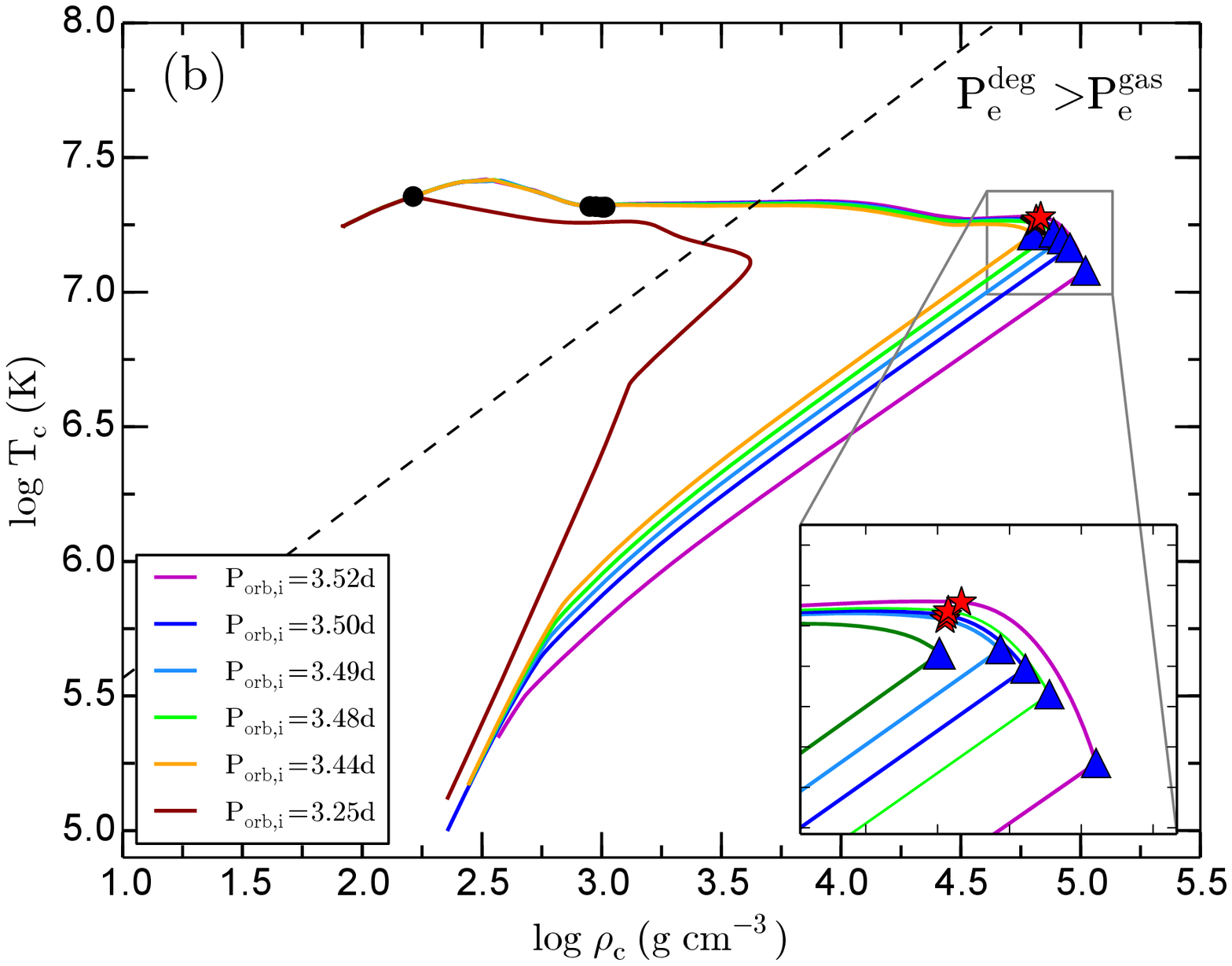}
\includegraphics[width=0.92\columnwidth,angle=0]{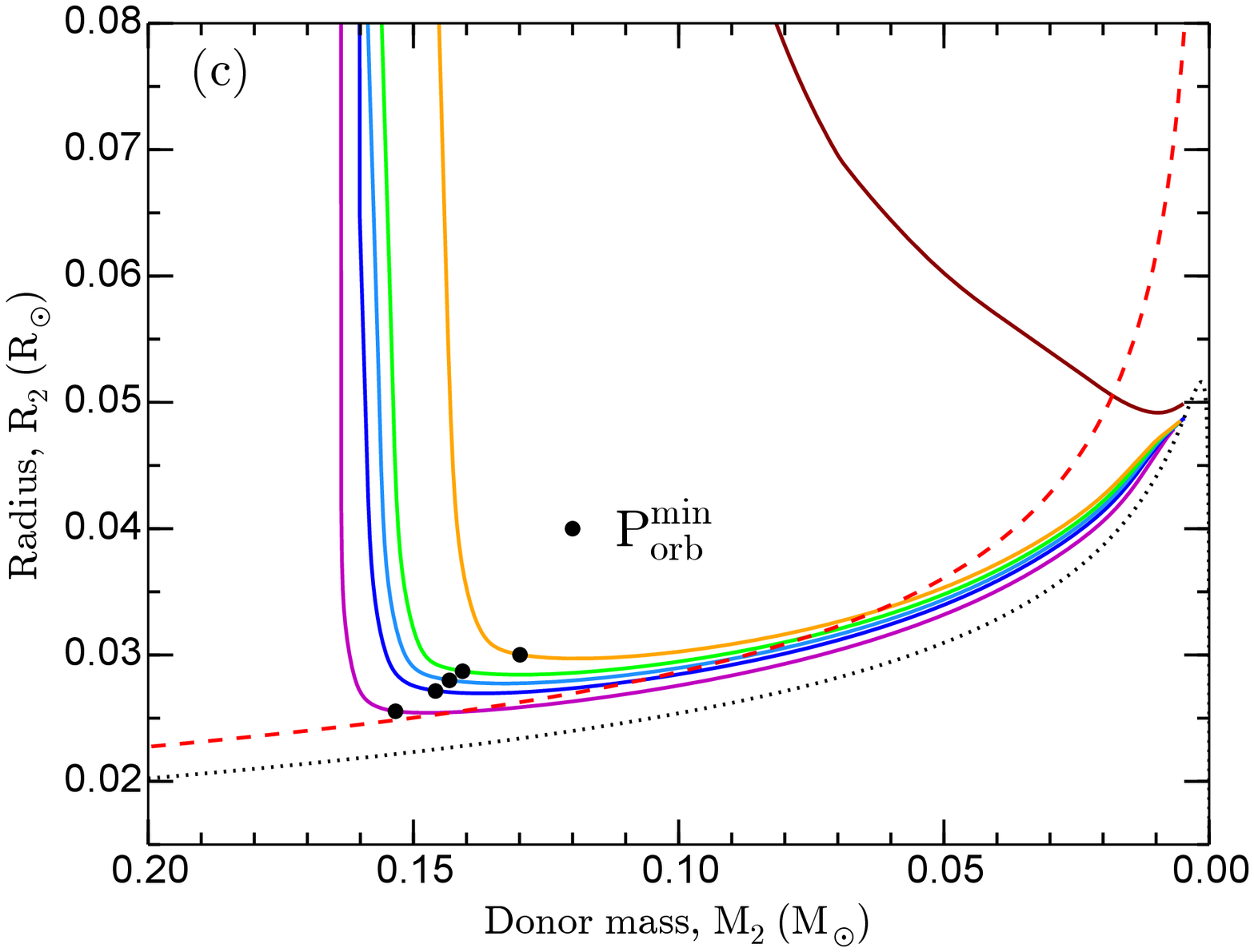}
\includegraphics[width=0.92\columnwidth,angle=0]{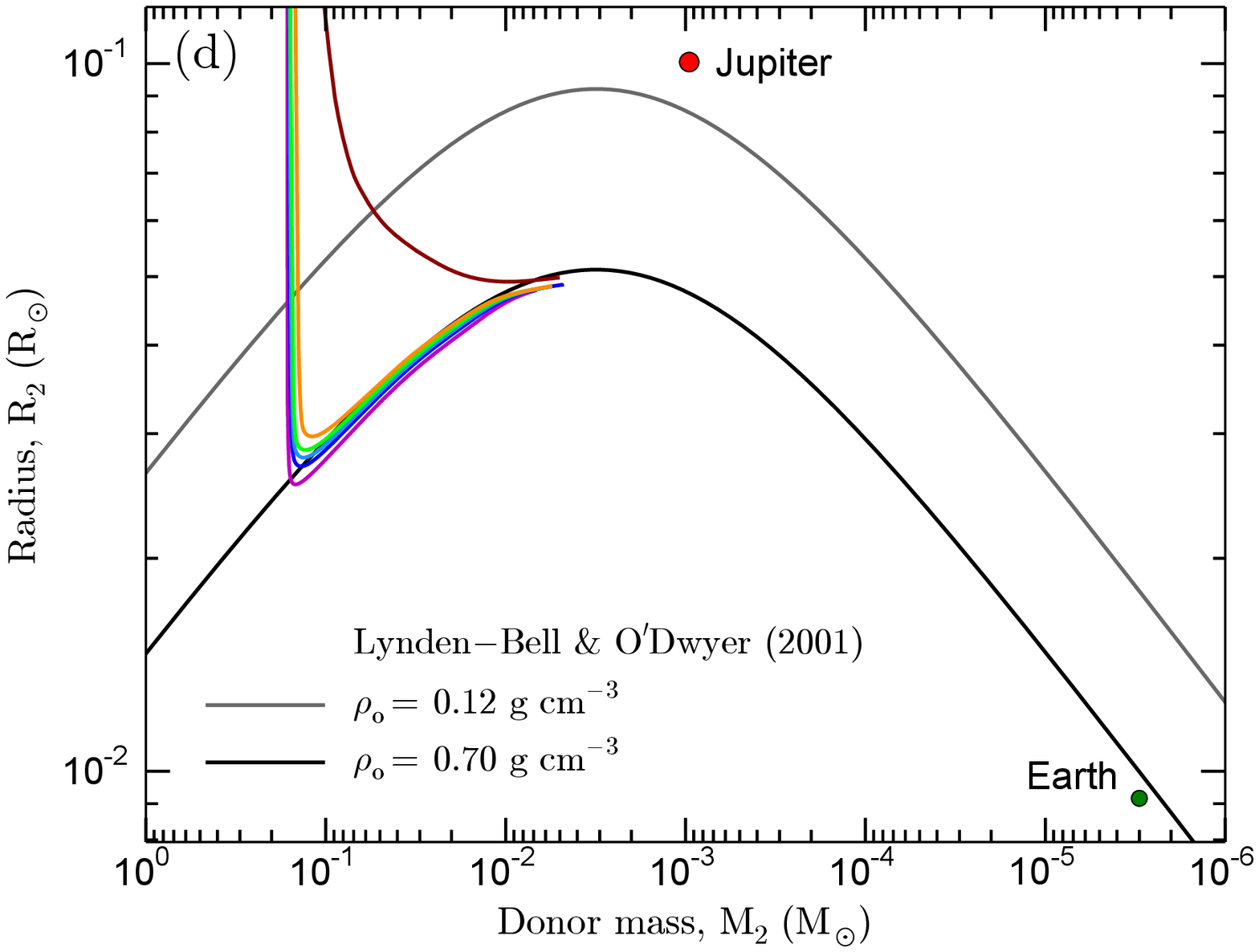}
\caption{Detailed properties of UCXB donor stars. The four plots displayed are as follows: HR--diagram with stellar ages for $P_{\rm orb,i}=3.48\;{\rm days}$ (upper-left panel,~a), central temperature vs. central mass density (upper-right panel,~b), stellar radius vs. mass (lower panels,~c and d). Common symbols are: black circles (onset LMXB RLO), red stars (termination LMXB RLO) and blue triangles (onset UCXB RLO). The dashed line in panel~b separates non-degenerate matter from degenerate matter. The red dashed and black dotted lines in panel~c are theoretical mass-radius relations: $R_{\rm WD}=0.013\;R_{\odot}\cdot (M_{\rm WD}/M_{\odot})^{-1/3}$ and that of Eggleton used in \citet{rjn87}. The two $M-R$ tracks plotted in panel~d (grey and black curves) are scaled from eq.~(34) in \citet{lo01} for cold planets.}
\label{fig:donor}
\end{center}
\end{figure*}

\section{Comparison to observations}\label{sec:obs}
At first sight in Fig.~\ref{fig:tracks}, we notice that our UCXB tracks can explain the location of the observed systems quite well. The data were taken from \citet{hie+13} and the error bars on $|\dot{M}_2|$ reflect bolometric correction uncertainty from the observed X-ray luminosity. 
The seven persistent and short-orbital period UCXBs ($P_{\rm orb}\simeq 10-25\;{\rm min}$) are either located on the ascending or the descending branch. A simple way to distinguish between the two options is the sign of $\dot{P}_{\rm orb}$. Unfortunately, the intrinsic value is difficult to derive in practice since many of the UCXBs are located in dense globular clusters, thereby suffering from acceleration in the cluster potential or gravitational perturbations from a nearby object. Moreover, a CB disc may cause $\dot{P}_{\rm orb}<0$ on the declining branch \citep{jcl17}. Statistically, however, UCXBs are much more likely to be on the declining branch since the temporal evolution is a lot slower along this branch \citep{db03,nyvt10}.

The three persistent sources with $P_{\rm orb}\simeq 40-50\;{\rm min}$ are best described by an LMXB system which evolves continuously into an UCXB without forming a detached WD (cf. red track for $P_{\rm orb,i}=3.25\;{\rm days}$). Alternatively, for a WD donor, a CB disc model can also account for these sources as a result of a significant increase in $|\dot{M}_2|$ \citep{hie+13}.

All the five transient systems likely populate the declining branch of the evolutionary tracks. Due to their wider orbits, the radius of their accretion disc is larger and its temperature lower, which causes thermal viscous instabilities and thus a transient behavior \citep{ldk08}. \citet{hie+13} suggested that this might be the reason why so relatively few UCXBs are seen in wide orbits with $P_{\rm orb}>30\;{\rm min}$ (keeping in mind that UCXBs should accumulate in wide orbits over time). The transient behavior allows radio MSPs to turn on, whereby the ``radio ejection mechanism" \citep{bdb02} can prevent further accretion. However, pulsar-wind irradiation of the donor may operate, possibly until $M_2 < 0.004\;M_{\odot}$ (if beaming is favourable), at which point the star is likely to undergo tidal disruption \citep{rs85} \citep[however, see also][]{pv88}; potentially leaving behind pulsar planets \citep{mlp16}. Our evolutionary tracks are terminated just before that when $M_2\simeq 0.005\;M_{\odot}$ and $P_{\rm orb}=70-80\;{\rm min}$.

To model even wider UCXBs with He~WD donors, will probably require irradiation effects to be included. Whereas these effects apparently have little effect on the UCXB evolution in an ($M_2,\,P_{\rm orb}$)--diagram \citep{bdh12}, they do accelerate the evolution of these systems \citep{vnvj12b}, which is needed to understand some of the observed relatively wide-orbit (post-UCXB) MSP binaries with very small companion star masses.

So far, we have not discussed the chemical composition of the observed spectra of the UCXB systems which holds the key to understanding their origin \citep{nyvt10}.
Although the nature of the donor stars has only been established firmly in some cases,  it seems already now quite clear that a variety of progenitor models is needed to explain their origin.
Our modelling presented here can account for the UCXBs with helium (or hydrogen) lines in their spectra.

\section{Summary}\label{sec:summary}
We have used MESA to calculate the complete  evolution of close binary systems leading to the formation and evolution of UCXBs. This includes numerical calculations (to our knowledge, for the first time) of stable mass transfer from a WD to an accreting NS.
In this work, we have concentrated on an initial binary with a relatively massive ZAMS donor star of $1.4\;M_{\odot}$. This allows for producing UCXBs at $P_{\rm orb}^{\rm min}$ within 10~Gyr. In another recent work, even more massive donor stars have been suggested to produce UCXBs \citep{cp16}. 
We also performed additional modelling using $M_2=1.2\;M_{\odot}$ and $\beta=0.3$. The evolutionary tracks of these systems closely resemble those plotted in  Fig.~\ref{fig:tracks}, although in this case no UCXBs are produced before a total age of 11.6~Gyr. A particular uncertainty in the first part of our calculations is the modelling of magnetic braking. It is evident that Nature produces tight LMXBs, CVs and NS--WD binaries via this channel, but the calibration of the effect as well as e.g. the required depth of the convective envelope remains uncertain \citep{itl14}.

To explain the full population of UCXBs, one needs to perform similar computations leading to donor stars evolved to different degrees of nuclear burning and which therefore have different chemical compositions (i.e. hydrogen-rich dwarf stars, naked helium stars, He~WDs or CO~WDs).
Whether such computations (with or without a common-envelope phase) are also possible for CO~WDs (or hybrid-CO~WDs) remains to be shown.

\section*{Acknowledgements}
RS thanks AIfA, University of Bonn, for funding during this MSc project and Pablo Marchant for help with MESA. We thank Craig Heinke, John Antoniadis and the referee, Lennart van Haaften, for very useful comments.

\vspace{0.5cm}
\bibliographystyle{mnras}
\bibliography{rahul_refs} 

\bsp	
\label{lastpage}
\end{document}